# Solderable Microcontroller-Integrated E-Textiles using UV-Tape-Assisted Laser Patterning Technique

## Author Information

### Affiliations

Naoto Tomita[1,3], Suguru Sato[1,3※], Toshihiro Takeshita[2], Aki Furusawa[2], Jarred Fastier-Wooller[1], Shun Muramatsu[1], Toshihiro Itoh[1], Michitaka Yamamoto[1※]

1 Department of Precision Engineering, Graduate School of Engineering, The University of Tokyo, Tokyo, Japan

2 National Institute of Advanced Industrial Science and Technology (AIST), Tsukuba City, Ibaraki, Japan

3 These authors contributed equally: Naoto Tomita, Suguru Sato

※email; sugurusato-hem@g.ecc.u-tokyo.ac.jp, yamamoto-michitaka@g.ecc.u-tokyo.ac.jp

### Contributions

N. T., S. S. and M. Y. conceived the idea, constructed the research frame, designed the experiment, and analyzed data. S. S. and M. Y. performed the experiments and wrote the original draft. S. S. designed and fabricated the prototype device and performed measurements using the fabricated device. T. T., T. I., and M. Y. performed supervision and funding acquisition. N. T., S. S., T. T., A. F., J. F., S. M., T. I., and M. Y. discussed results and reviewed the manuscript.

## Abstract

In this study, we developed a UV-tape-assisted laser patterning (UT-Laser) technique that enables the simple transfer-based formation of wiring with line widths below 200 μm onto textile substrates. With the rapid advancement of wearable devices capable of acquiring various types of physiological and environmental information, research on electronic textiles (e-textiles)—in which electronic components are integrated into fabrics and clothing—has progressed considerably. However, integrating high-performance, rigid electronic components onto textiles remains challenging: the diameter of textile fibers limits the formation of fine wiring, making reliable mounting of such components difficult. To address these challenges, we devised the UT-Laser technique, in which thin foil or film materials are laser vector-cut on UV tape, and the adhesive strength is controlled through UV exposure. The unnecessary portions are selectively and collectively peeled away to form fine wiring, which is subsequently transferred onto the textile substrate. This approach enables facile fabrication of fine wiring with line widths below 200 μm on textiles. Furthermore, by forming fine wiring from a flexible copper clad laminate and transferring it onto heat-resistant glass cloth, electronic components can be soldered directly, allowing the fabrication of e-textile devices capable of withstanding more than 10,000 bending cycles. The prototype e-textile device fabricated using the proposed method integrates a microcontroller, USB connector, battery holder, flash memory, inertial measurement unit, and environmental sensors, and successfully acquires data related to stair climbing, respiration, and changes in body temperature during sleep.

## Introduction

Wearable electronics have attracted considerable attention for their potential applications in areas such as health monitoring of humans and animals[1,2], behavioral analysis and motion control[3], and human–machine interfaces[4,5]. Notably, highly functional wearable devices have emerged that integrate sensors, actuators, memory, and signal-processing circuits onto flexible substrates[6,7], including devices fabricated entirely on flexible materials such as polyimide (PI)[8], as well as flexible hybrid electronics that incorporate high-performance rigid components, such as silicon integrated circuits, onto flexible substrates[9]. More recently, new approaches, including stretchable electronics[10], electronic skin (e-skin) devices[11], and electronic textiles (e-textiles)[12], have been developed to enable more comfortable and stable measurements. Among these, e-textile devices are particularly promising as next-generation wearable technologies because they use textiles that are continuously worn, biocompatible with human skin, and highly breathable[12].

E-textile devices can be broadly classified into two approaches: directly imparting functionality to fibers or textiles[13], and forming wiring on textiles to integrate rigid electronic components, such as silicon integrated circuits[14]. The former approach includes examples such as fabricating transistors by coating fibers with functional materials[15], integrating bioelectrodes onto textiles through organic conductor coatings[16], and creating chemical sensors via polyaniline coating[17]. Although these methods effectively exploit intrinsic textile properties such as flexibility and breathability, their performance as electronic devices remains inferior to that of rigid electronic components, particularly silicon integrated circuits. Thus, the approach of forming wiring on textiles and integrating rigid electronic components is currently regarded as more promising in terms of achieving higher performance and multifunctionality[14].

For the integration of rigid electronic components onto textiles, early approaches include connecting flexible printed circuits using conductive threads[18] and implementing LEDs on ribbons with copper wires[19]. More recently, techniques based on fiber plating have been proposed to form wiring capable of supporting more complex circuits. For example, copper-plated wiring with a 1-mm pitch has been used to integrate ultrasonic elements for fabricating ultrasound probes[20]. Other studies have demonstrated laser processing of metallized textiles to create multilayer structures by interconnecting wiring with liquid metals, as well as the integration of microchips or flexible sensors with minimum pitches of approximately 0.8 mm[21]. In addition, in-textile photolithography has been used to form fine wiring and integrate devices such as sweat sensors and ESP32 modules[22]. Despite these advances, plating-based methods continue to face several challenges, including limitations on wiring width imposed by the textile fiber structure, the requirement for chemical processing and large-scale equipment, and variations in electrical resistance caused by fiber movement[20-22].

As an alternative approach, techniques have been proposed that use metal foils adhered to textiles to serve as wiring[23,24]. Because these methods use continuous metal foils, they are not limited by the diameter of textile fibers and can provide stable electrical interconnects. For example, copper foil attached to fabric has been selectively cut by laser processing, followed by manual removal of unnecessary regions, to fabricate high-frequency antennas with line widths as small as 0.3 mm[23]. More recently, xurography has been applied to wiring formation on textiles, in which metal foil is patterned using a cutting plotter and subsequently transferred onto the textile. This approach enabled the fabrication of ultrasonic probes with ultrasonic elements arranged at a 600-μm pitch and demonstrated stable electrical resistance during repeated bending-cycle tests[24]. However, these techniques have thus far been applied mainly to relatively simple geometries and low-density patterns, and considerable challenges remain in realizing miniaturized, complex, and high-density wiring. A key limitation of previous xurography-based approaches is the processing accuracy achievable with commercially available cutting plotters. In addition, through-cut and weeding processes, such as those used in xurography or laser vector cutting followed by removal of unnecessary regions, require manual operations with tweezers during the weeding step[25,26]. As wiring patterns become finer and more complex, the associated workload increases substantially, even though vector-based processing offers considerably shorter cutting times than raster-based methods[27]. Therefore, there is a strong need for a novel fabrication process that enables fine, intricate circuit wiring while eliminating the need for labor-intensive and complex manual weeding procedures.

To address these challenges, we developed a novel UV-tape-assisted laser patterning (UT-Laser) technique that enables the facile fabrication of complex, fine circuit patterns on textiles. This technique involves laser vector cutting of foil or film materials laminated onto UV tape, followed by partial UV exposure to locally control the in-plane adhesive strength of the tape. This controlled adhesion allows unnecessary regions to be removed collectively, thereby greatly simplifying the cleaning and weeding processes required in conventional vector-based fabrication. The UT-Laser technique is applicable to a wide range of laser-processable foils and films; using this method, we successfully patterned and transferred copper foil, aluminum foil, flexible copper clad laminate (FCCL), and PI film into fine wiring with line widths below 200 μm. Using this approach, we also realized multilayer structures comprising two or three layers. Furthermore, to prototype highly durable devices, we selected material combinations that enabled the transfer of FCCL wiring onto heat-resistant glass cloth, allowing electronic components to be soldered directly. This configuration resulted in a robust e-textile device structure capable of withstanding more than 10,000 bending cycles. Using the UT-Laser technique, we successfully prototyped e-textile devices integrating a microcontroller, USB connector, battery holder, flash memory, inertial measurement unit (IMU), and environmental sensor, thereby demonstrating the practicality and versatility of the proposed method.

## Results

### Proposed Concept and Fabrication Process

Figure 1 shows the device's conceptual structure and the core UT-Laser technique. The architecture features fine-patterned interconnects fabricated from FCCL on a glass-fabric substrate, with electronic components—particularly a packaged microcontroller with fine-pitch interconnects—soldered onto the substrate. The key aspect of the fabrication process is the UT-Laser technique: thin foil or film materials are placed on UV tape, laser vector-cut, and unnecessary regions are collectively removed after partial UV exposure of the tape to control the in-plane adhesive strength. This UT-Laser technique enables fine patterning and transfer of a wide variety of metal foils and films. Although a broad range of textile substrates can be selected for transfer, glass cloth was chosen in this study. When copper-based materials are patterned and transferred onto glass cloth, the resulting material composition becomes equivalent to that of conventional printed circuit boards (PCBs). Thus, standard PCB manufacturing processes—such as solder mask printing, automated chip placement, and solder reflow—can be directly applied, enabling the use of existing manufacturing infrastructure. This approach allows wiring pitches and mechanical durability comparable to those of conventional PCBs, while simultaneously imparting textile characteristics such as flexibility and breathability.

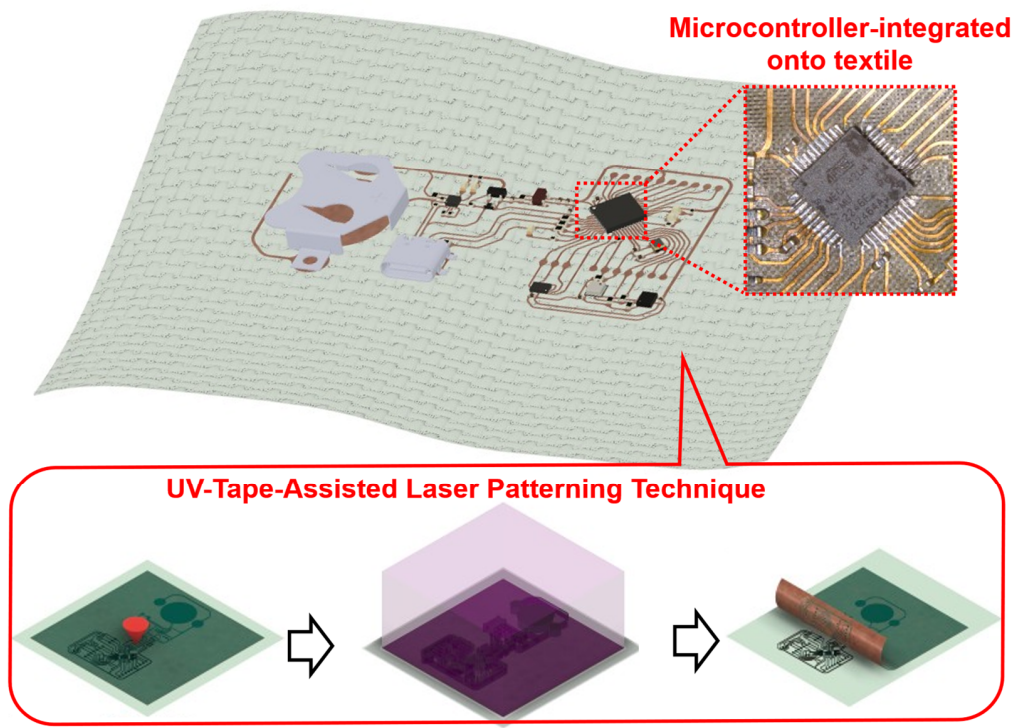

Figure 2a shows the overall fabrication process of the device developed in this study. The process can be broadly divided into three main steps: (i) the UT-Laser step for forming fine wiring on UV tape (Fig. 2a-I), (ii) the transfer step for transferring the patterned wiring from the UV tape onto the textile substrate (Fig. 2a-II), and (iii) the mounting step, in which rigid electronic components are mounted and electrically connected (Fig. 2a-III).

In the UT-Laser step, a foil or film material laminated with a thermosetting adhesive film is attached to UV tape and laser vector-cut into the desired wiring pattern (Fig. 2a-i). After laser cutting, UV light is selectively irradiated onto regions other than the wiring using a printed OHP film mask (Fig. 2a-ii). The unnecessary portions are then collectively peeled away from the tape (Fig. 2a-iii, iv). The peeling process is shown in Fig. 2b, and a video demonstration is provided in Supplementary Video 1.

In the transfer step (Fig. 2a-II), the fine-patterned wiring on the UV tape is further exposed to UV light to reduce its adhesion to the tape, and is then placed onto the textile substrate (Fig. 2a-v). The wiring is bonded to the textile by thermally curing the thermosetting adhesive film. Finally, only the UV tape is peeled away (Figs. 2a-vi, vii), and the adhesive layer is fully cured by annealing. Figure 2c and Supplementary Video 2 illustrate the process of peeling the UV tape from the textile. When a protective layer is required on top of the wiring, or when multilayer wiring is desired, this process can be repeated to fabricate two- or three-layer structures. It should be noted that increasing the number of layers results in increased stiffness of the patterned regions, which may compromise the inherent softness of textiles and should therefore be considered during device design.

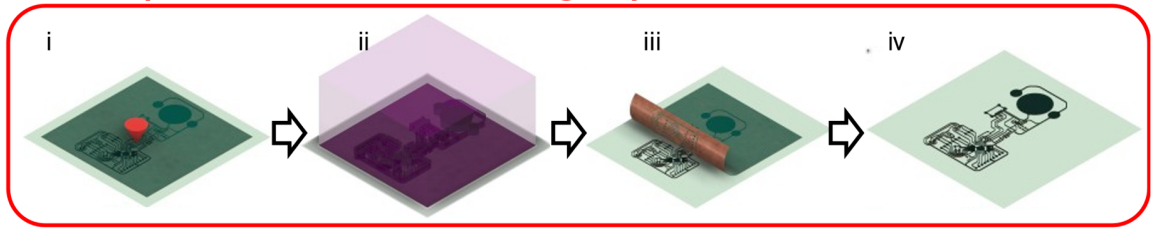

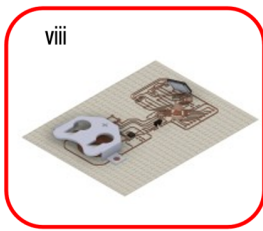 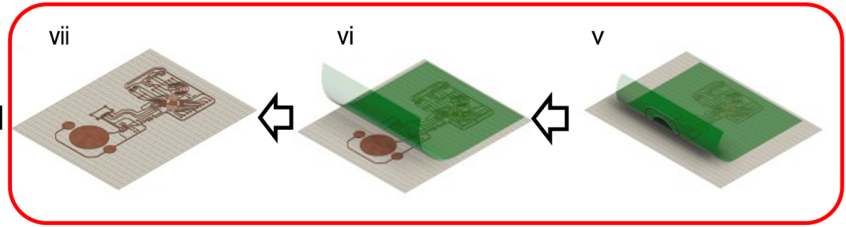

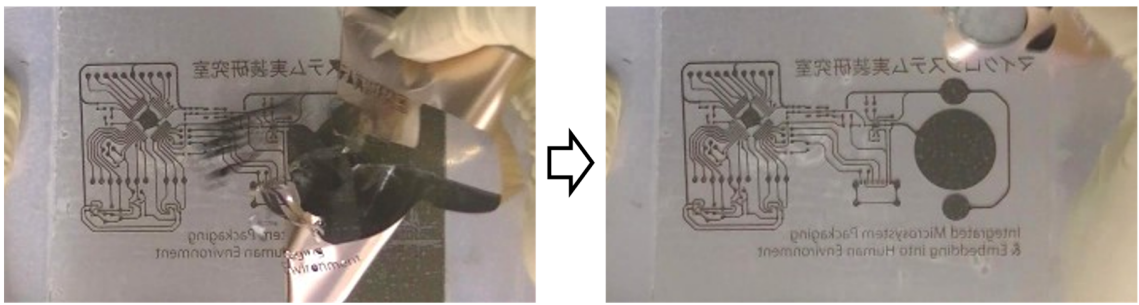

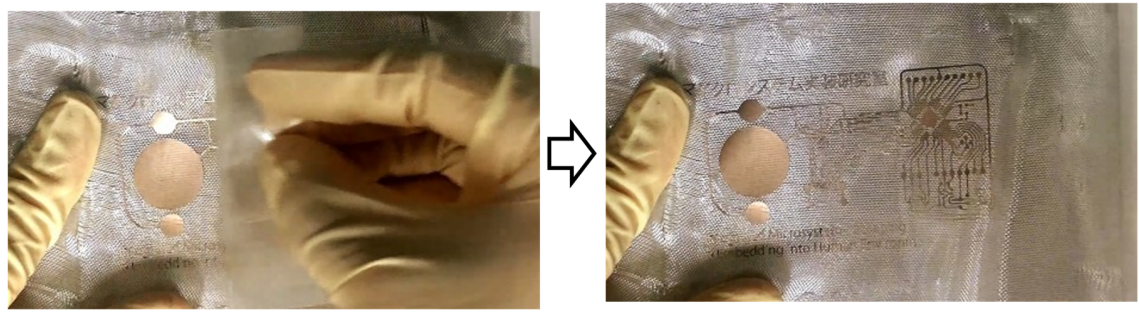

Fig. 2 **a:** Device fabrication process, divided into three main steps: UT-Laser patterning of fine wiring on UV tape [I], transfer of the wiring from UV tape onto the textile [II], and soldering of electronic components [III]. **b:** Example of collective removal of unnecessary portions during the UT-Laser step (corresponding to (a)-(iii)). **c:** Transfer of fine wiring from UV tape onto the fabric (corresponding to (a)-(vii)).

In the mounting step (Fig. 2a-III, or Fig. 2-viii), rigid electronic components are attached to the wiring. The optimal mounting method depends on the wiring type, the textile substrate, and the thermal resistance of the adhesive film. Specifically, using glass cloth as the substrate and copper-based materials such as FCCL for the wiring, standard PCB processes—including solder screen printing, chip placement with chip mounters, and solder reflow—can be employed, taking advantage of existing manufacturing infrastructure.

Evaluation Results of the Fabricated Wiring

Examples of fine-patterned wiring fabricated on glass cloth using aluminum and FCCL foils via the UT-Laser technique are shown in Figs. 3a and b, respectively. Supplementary Fig. 1 shows additional circuit patterns processed using this method, demonstrating successful patterning and transfer of wiring with a 500-μm pitch (300-μm line width and 200-μm spacing) onto textile substrates. Even finer features were also attainable; the straight-line patterns with a 150-µm line width (300-µm pitch) were successfully patterned as shown in Fig. 3c. Supplementary Fig. 2 further shows photographs of straight-line patterns with line widths of 200, 300, and 500 μm, respectively, while Supplementary Fig. 3 shows an example of a wiring pattern with a 200-μm line width. Notably, the UT-Laser technique is applicable not only to metal foils but also to a wide range of film materials; Fig. 3d shows an example in which the same pattern was fabricated using a PI film. Moreover, by repeating the UT-Laser and transfer processes, multilayer structures can be constructed. Figure 3e shows an example of a PI protective layer formed on FCCL wiring by repeating these steps twice. Figure 3f shows a multilayer structure formed by stacking aluminum, PI, and copper patterns through three repetitions of the UT-Laser and transfer steps. Details of the wiring pattern and the corresponding cross-sectional structure are shown in Supplementary Fig. 4. These results demonstrate that the proposed technique enables the fabrication of multilayer structures and interconnects.

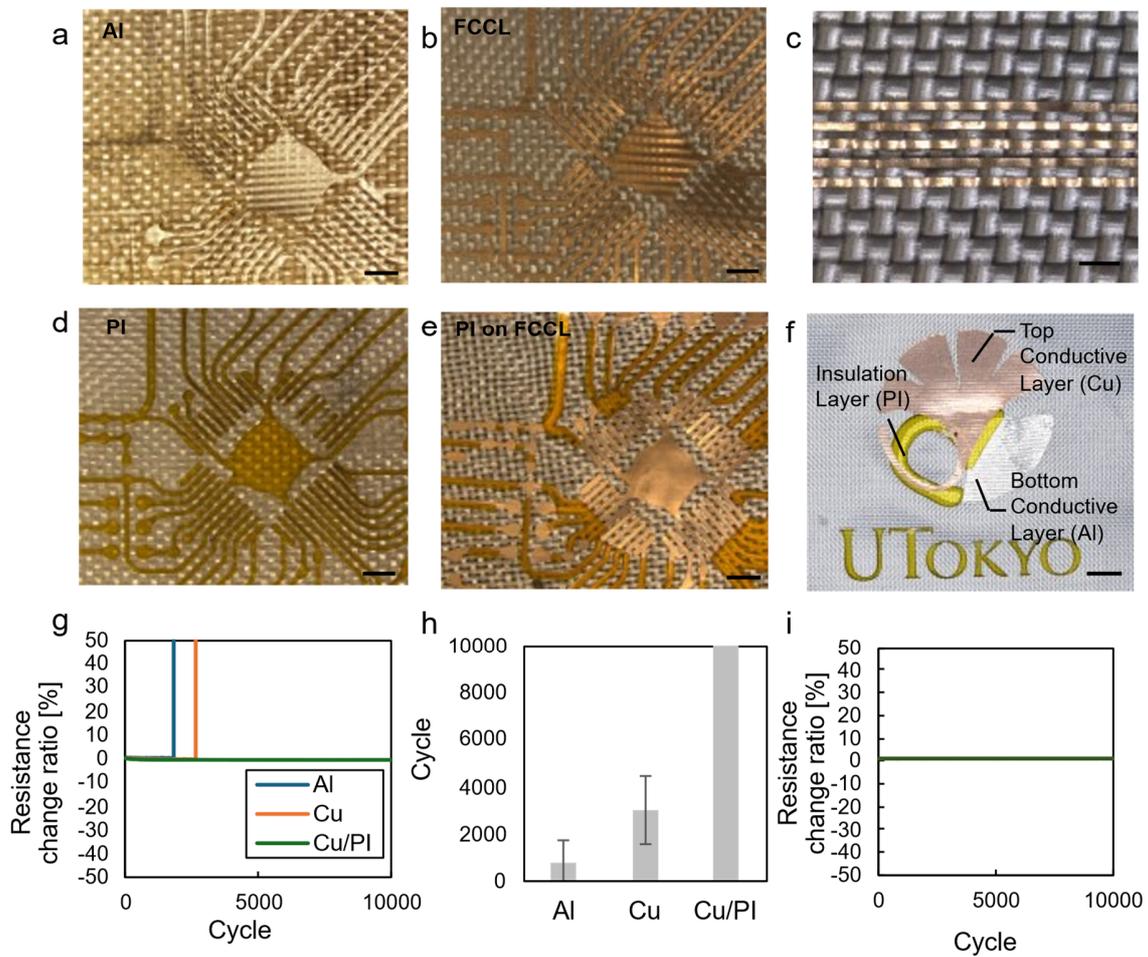

Fig. 3 **a–b:** Optical images of aluminum and FCCL wiring fabricated on glass cloth using the UT-Laser technique. Scale bar: 2 mm. **c:** Optical image of wiring with 150-μm line width and 150-μm spacing. Scale bar: 600 μm. **d:** Example of a patterned polyimide film on glass cloth. Scale bar: 2 mm. **e:** Polyimide protective layer formed on FCCL wiring by repeating the UT-Laser and transfer processes twice. Scale bar: 2 mm. **f:** Optical image of a multilayer wiring structure fabricated by repeating the UT-Laser and transfer processes three times. Scale bar: 10 mm. **g, h:** Bending-cycle test results up to 10,000 cycles for wiring on glass cloth, compared by wiring material. **i:** Typical bending-cycle test result up to 10,000 cycles for a sample with a chip mounted on FCCL wiring.

However, it should be noted that the current implementation does not support the formation of through-vias, requiring separate connections between exposed terminals. In addition, increasing the number of layers results in a thicker device, which may compromise the inherent softness of the textile substrate.

To identify the optimal wiring material for e-textile devices, we fabricated wiring with a width of 500 μm and a length of 100 mm from aluminum, copper, and FCCL foils on glass cloth and evaluated their bending-cycle durability up to 10,000 cycles using a sliding-plate test. The tests were performed with a curvature radius of 10 mm using a home-built apparatus. Representative samples and the test setup are shown in Supplementary Fig. 5. Typical results of the bending-cycle tests for each material are shown in Fig. 3g, indicating that the electrical resistance remains nearly constant until fracture occurs. This stability of resistance prior to failure is a key advantage of the UT-Laser technique, which uses foil-based wiring. The number of cycles to failure varied considerably among the materials, with durability increasing in the order of aluminum, copper, and FCCL. Figure 3h compares the average number of bending cycles to failure for each wiring material. While aluminum and copper wiring fractured after a limited number of cycles, FCCL wiring withstood more than 10,000 bending cycles without failure. This superior durability is attributed to the multilayer structure of FCCL, in which cracks initiated in the copper layer are effectively arrested by the underlying PI layer. Furthermore, to evaluate durability under practical device conditions, bending-cycle tests up to 10,000 cycles were performed on samples with 1608-size (1.6 mm × 0.8 mm) 0-$\Omega$ chip resistors mounted on FCCL wiring on glass cloth, as shown in Fig. 3i. Even with chip mounting, the samples endured over 10,000 bending cycles, demonstrating sufficient mechanical robustness for e-textile applications. It should be noted that the initial resistance of the FCCL-based wiring was approximately 0.4 $\Omega$.

## E-textile Device Fabrication

To demonstrate the utility of the UT-Laser technique, we fabricated a prototype e-textile device by transferring fine FCCL-based circuit patterns onto glass cloth and mounting electronic components. The prototype device integrates a microcontroller, USB connector, battery holder, flash memory, IMU, and environmental sensor. The device operates on battery power, records sensor data to flash memory, and enables data retrieval via a USB interface. The fabricated circuit pattern is shown in Supplementary Fig. 1, while the circuit schematic and the list of components used are provided in Supplementary Fig. 6 and Table 1, respectively. An optical image of the completed device is shown in Fig. 4a. To allow for future expansion,

I/O ports were also implemented, enabling additional sensors to be connected by soldering. The finest wiring pitch in the device is associated with the 44-pin VQFN-packaged microcontroller, which requires a wiring pitch of 500 µm. As shown in Fig. 4b, sufficiently fine wiring was successfully fabricated and solder-mounted. Solder paste was applied by mask printing, components were placed using a desktop mounter (ALMAS 8, ePRONICS Co., Ltd.), and soldering was completed by reflow on a hot plate. Figure 4c shows the flexibility characteristic of the textile-based device.

The prototype e-textile device was attached to the inner surface of clothing, and various movements were monitored using the integrated IMU and environmental sensor. Figure 4d shows representative results obtained by mounting the device on a garment and recording changes in acceleration and pressure during descent of a three-floor staircase. The data clearly show pronounced acceleration during stair descent, accompanied by corresponding pressure variations. As an example of the gas-sensing capability provided by the environmental sensor, the device was placed inside a mask and exposed to exhaled breath containing alcohol; the resulting sensor response is shown in Fig. 4e. The breathability of the e-textile enables a clear signal that closely follows the exhalation cycle. Furthermore, after showering and changing clothes, the device was worn on clothing, and data were collected during sleep, as shown in Fig. 4f. The device recorded changes in temperature and humidity inside clothing during sleep. Thus, the capability to monitor humidity variations during sleep highlights a key advantage of soft and breathable e-textiles.

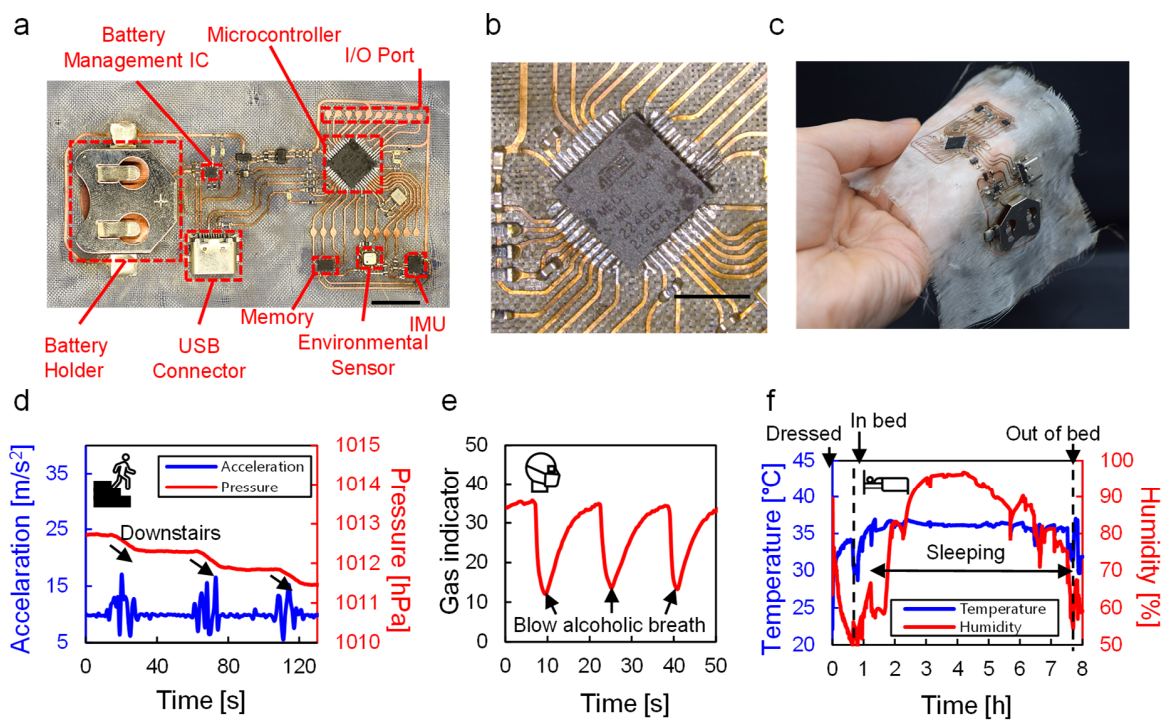

Fig. 4 **a:** Optical image of the fabricated device. Scale bar: 10 mm. **b:** Enlarged view of the microcontroller section, showing high-precision mounting of a microcontroller with a 500-µm pitch. Scale bar: 5 mm. **c:** Photograph illustrating the textile-like flexibility of the device. **d:** Measurement results when the device was worn inside clothing, recording acceleration and pressure changes while descending three floors of stairs. **e:** Example of the gas sensor response when the device was placed inside a mask and exposed to breath containing alcohol. **f:** Temperature and humidity measurements inside clothing during sleep, collected after the device was attached after showering and changing.

## Discussion

In this study, we developed a novel UT-Laser technique that enables the facile formation of circuit patterns on UV tape and their subsequent transfer onto textiles through controlled modulation of the tape's in-plane adhesive properties. Using this method, we successfully processed and transferred materials such as aluminum, copper, and FCCL into fine wiring patterns with line widths below 200 μm, and demonstrated that multilayer wiring structures can be realized by repeating the UT-Laser process. Furthermore, by forming copper-based wiring on glass cloth—a material widely used in conventional PCB substrates—and solder-mounting rigid electronic components, we established a robust e-textile device architecture suitable for practical applications.

In practice, we fabricated durable e-textile devices by forming fine FCCL-based wiring on glass cloth and solder-mounting electronic components. The utility of both the devices and the UT-Laser technique was demonstrated by successfully capturing daily activity and physiological data. These results confirm the effectiveness and versatility of the proposed approach for future e-textile and wearable device applications.

## Methods

### Materials

The thin materials used in this study included 11-μm-thick aluminum foil (ALH30×50, MonotaRO Co., Ltd., Osaka, Japan), 10-μm-thick copper foil (C1020R-H, Irie Shokai Co., Ltd., Tokyo, Japan), and a thermoplastic PI film with a thickness of 25 μm (Cat. No. 3-8010-01, AS ONE Corporation, Osaka, Japan). The FCCL consisted of Cu/PI layers with thicknesses of 9/12.5 μm (PNS H0509RAC, Arisawa Manufacturing Co., Ltd., Joetsu, Niigata, Japan). The adhesive (bonding) sheet used for transfer was AU-15 (Arisawa Manufacturing Co., Ltd., Joetsu, Niigata, Japan). The UV tape used was from the Adwill series (LINTEC Corporation, Tokyo, Japan), and the glass cloth substrate was 1080NT-127 S640S (Arisawa Manufacturing Co., Ltd., Joetsu, Niigata, Japan). The solder used for electronic component mounting was M705-ULT369 Type 5 (Senju Metal Industry Co., Ltd., Tokyo, Japan).

### Fabrication Equipment

Laser processing was performed using a UV laser cutting machine. Photomasks were fabricated by printing onto OHP transparency film (VF-1, KOKUYO Co., Ltd., Osaka, Japan), and UV exposure was performed using a UV exposure light box (BOX-S3000, Sunhayato Corp., Tokyo, Japan). Thermal bonding and curing were performed using a rosin heat press machine (WXQ; 600 W, 110 V; purchased from Amazon.co.jp, ASIN B0D6KGJ4F). Electronic component placement was performed using a desktop chip mounter (ALMAS 8; ePRONICS Co., Ltd., Tokyo, Japan).

### Fabrication Process

The laser-cutting conditions were optimized for each material. After laser processing, UV light was applied for 2 min, after which the unnecessary portions were removed. During the transfer process, UV light was irradiated for more than 2 min, followed by thermal bonding at 110°C under a load of 1 ton for 1 min to adhere the wiring to the glass cloth. Subsequently, only the UV tape was peeled off. To fully cure the adhesive layer, the samples were heated at 160°C for 1 h. For electronic component mounting, solder paste was printed using a handmade metal mask, components were placed using the chip mounter, and soldering was completed by reflow on a hot plate at approximately 300°C.

Evaluation Setup

Bending-cycle tests were performed using a custom-built sliding-plate test apparatus with a curvature radius of 10 mm and a stroke length of 85 mm. The test speed was set to 60 rpm for wiring samples and 120 rpm for samples with chips mounted on FCCL wiring. During testing, the wiring was connected to resistance-measurement instruments, including the Impedance Analyzer function in WaveForms, an Analog Discovery 3 (Digilent, Pullman, WA, USA), or a digital multimeter (Model 34465A, Keysight Technologies, Santa Rosa, CA, USA).

## Acknowledgements

This work was supported in part by Grant-in-Aid from the Foundation for Technology Promotion of Electronic Circuit Board, the Ozawa and Yoshikawa Memorial Electronics Research Foundation. We would like to thank LINTEC Corporation and Arisawa Manufacturing Co., Ltd. for kindly providing the samples used in this study. The authors would like to greatly thank Prof. Takamatsu at State University of New York, Binghamton, the previous supervisor of Naoto Tomita and Suguru Sato.

# Ethics Declarations

Competing interests

The authors declare no competing interests.

# Supplementary Information

Supplementary Information

    Supplementary Figs. 1–6 and Table 1.

Supplementary Video 1

    The video of the peeling process.

Supplementary Video 2

    The video of the transfer process to the clothes.